# Parallel Computing Based Solution for Reliability-Constrained Distribution Network Planning

Yaqi Sun, Wenchuan Wu, *Fellow, IEEE*, Yi Lin, Hai Huang, Hao Chen

*Abstract*—The main goal of distribution network (DN) expansion planning is essentially to achieve minimal investment constrained with specified reliability requirements. The reliability-constrained distribution network planning (RcDNP) problem can be cast an instance of mixed-integer linear programming (MILP) which involves ultra-heavy computation burden especially for large scale DNs. In this paper, we propose a parallel computing-based acceleration algorithm for solve RcDNP problem. The RcDNP is decomposed into a backbone grid and several lateral grid problems with coordination. Then a parallelizable augmented Lagrangian algorithm with acceleration strategy is developed to solve the coordination planning problems. In this method, the lateral grid problems are solved in parallel through coordinating with the backbone grid planning problem. To address the presence of nonconvexity, Gauss–Seidel iteration is adopted on the convex hull of the feasible region constructed by the decomposition method. Under mild conditions, the optimality and convergence of this algorithm is proven. The numerical tests show the proposed method can significantly reduce the computation time and make the RcDNP applicable for real-world problems.

*Index Terms*—Distribution network, expansion planning, reliability, parallel computing

## NOMENCLATURE

### A. Sets and Vectors

| | |
|---|---|
| $\Lambda^C$ | Set of alternatives types of conductor. |
| $\Lambda^T$ | Set of alternative types of transformer. |
| $\Psi_B$ | Set of branches. |
| $\Psi_N$ | Set of nodes. |
| $\Psi_T$ | Set of transformers. |
| $\Psi_F$ | Set of feeders. |
| $\Psi_{Tr}^{out}$ | Set of transformer outlet branches. |
| $\Psi_{SS}$ | Set of substation nodes. |
| $\Psi_i$ | Set of nodes connected to node $i$. |
| $D_b, D_{s,n}$ | Subset of convex hull $conv(X_b)$ and $conv(X_{s,n})$. |
| $w_b, w_{s,n}$ | Vector of Lagrangian multipliers. |
| $\hat{w}_b, \hat{w}_{s,n}$ | Auxiliary vector for updating Lagrangian multipliers. |
| $x_b, x_{s,n}$ | Vector of decision variables of backbone grid and the $n$th sub-area planning model. |
| $\hat{x}_b, \hat{x}_{s,n}$ | Auxiliary vector to expand subset $D_b$ and $D_{s,n}$. |
| $X_b, X_{s,n}$ | Feasible set of backbone grid and the $n$th sub-area planning model. |
| $z_b, z_{s,n}$ | Vector of coordination variables describing boundary conditions for backbone grid and $n$th sub-area planning model. |
| $(\cdot)^k$ | Variables in the $k$th iteration. |

### B. Parameters

| | |
|---|---|
| $\tau_{xy}^{SW}$ | Duration of the switching-only interruptions associated with the failure of the branch connecting nodes $x$ and $y$. |
| $\tau_{xy}^{RP}$ | Duration of the repair-and-switching interruptions associated with the failure of the branch connecting nodes $x$ and $y$. |
| $\varepsilon_{SAIDI}$ | SAIDI requirement of system. |
| $d_{ij}$ | Distance between node $i$ and node $j$. |
| $k_{max}$ | Maximum number of iterations of the main loop. |
| $l_{ij}^{a,0}$ | Equal to 1 when type $a$ conductor exists at branch $ij$ originally. |
| $m_{Tr}^{a,0}$ | Equal to 1 when type $a$ transformer exists at transformer $Tr$ originally. |
| $n_s^0$ | Equal to 1 when a substation exists at node $s$ originally. |
| $n_{sub}$ | Number of sub-areas. |
| $N_s$ | Maximum number of transformers in the substation. |
| $NC_i$ | Number of customers at node $i$. |
| $r_C^a$ | Resistance unit of type $a$ conductor. |
| $S_C^a$ | Capacity of type $a$ conductor. |
| $S_T^a$ | Capacity of type $a$ transformer. |
| $T_{max}$ | Maximum number of iterations of the inner loop. |
| $x_C^a$ | Inductance unit of type $a$ conductor. |
| $\overline{(\cdot)}/\underline{(\cdot)}$ | Upper and lower bounds of variable $(\cdot)$. |

### C. Continuous Variables

| | |
|---|---|
| $\underline{\phi}_b, \underline{\phi}_{s,n}$ | Value of dual function associated with backbone |

Manuscript received XX, 2022. This work was supported in part by the State Grid Science and Technology Program of China (Grant. 5100-202121561A-0-5-SF).

Y.Sun(sun-yq20@mails.tsinghua.edu.cn) and W. Wu(Corresponding Author, e-mail: wuwench@tsinghua.edu.cn) are with the State Key Laboratory of Power Systems, Department of Electrical Engineering, Tsinghua University, Beijing 100084, China and Sichuan Energy Internet Research Institute,Tsinghua University. Y. Lin, H. Huang and H. Chen are with State Grid Fujian Electric Power CO. LTD.

| | grid and $n$th sub-area planning model. |
|---|---|
| $\hat{\phi}_b, \hat{\phi}_{s,n}$ | Value of cutting plane function used to approximate the dual function associated with backbone grid and $n$th sub-area planning model. |
| $CIF_i^t$ | Customer interruption frequency of node $i$ at stage $t$. |
| $CID_i^t$ | Customer interruption duration of node $i$ at stage $t$. |
| $h_i^{f,t}$ | $h_i^f \in [0,1]$ and equal to 1 when node $i$ is supplied by feeder $f$ under normal operating condition at stage $t$. |
| $h_{ij}^{f,t}$ | $h_{ij}^f \in [0,1]$ and equal to 1 when branch $ij$ is supplied by feeder $f$ under normal operating condition at stage $t$. |
| $n_i^{xy,t}$ | Substitution variable of product term $\lambda_{xy}^t \tau_{xy}^t p_i^{xy,t}$ for equivalent outlet branch of equivalent distribution source in the sub-area at stage $t$. |
| $P_i^t, Q_i^t$ | Active and reactive load demand of node $i$ at stage $t$. |
| $P_i^{xy,t}, Q_i^{xy,t}$ | Active and reactive demand at node $i$ after post-fault network reconfiguration due to a fault at branch $xy$ (or under normal operation condition if $xy$ = NO) at stage $t$. |
| $P_{ij}^{xy,t}, Q_{ij}^{xy,t}$ | Active and reactive power flow through branch $ij$ (from node $i$ to node j) after post-fault network reconfiguration due to a fault at branch $xy$ (under normal operation condition if $xy$ = NO) at stage $t$. |
| $P_{Tr}^{xy,t}, Q_{Tr}^{xy,t}$ | Active and reactive power flow through transformer $Tr$ after post-fault network reconfiguration due to a fault at branch $xy$ (under normal operation condition if $xy$ = NO) at stage $t$. |
| $SAIDI^t$ | System Average Interruption Duration Index at stage $t$. |
| $U_i^{ss,t}$ | Square of the voltage at reference node $i$ at stage $t$. |
| $U_i^{xy,t}$ | Square of the voltage at node $i$ due to a fault at branch $xy$ (or under normal operation condition if $xy$ = NO) at stage $t$. |

### D. Binary variable

| | |
|---|---|
| $l_{ij}^{a,t}$ | Equal to 1 when type $a$ conductor is installed at branch $ij$, otherwise equal to 0 at stage $t$. |
| $l_{ij}^t$ | Equal to 1 when there exists a conductor at branch $ij$, otherwise equal to 0 at stage $t$. |
| $m_{Tr}^{a,t}$ | Equal to 1 when type $a$ transformer is installed at transformer $Tr$, otherwise equal to 0 at stage $t$. |
| $m_{Tr}^t$ | Equal to 1 when transformer $Tr$ exists, otherwise equal to 0 at stage $t$. |
| $n_s^t$ | Equal to 1 when a substation is newly constructed at node $s$, otherwise equal to 0 at stage $t$. |
| $p_i^{xy,t}$ | Equal to 1 when node $i$ is affected by the outage due to a fault in branch $xy$ at stage $t$. |
| $q_i^{xy,t}$ | Equal to 1 when node $i$ is still in outage after network reconfiguration following a fault in branch $xy$ at stage $t$. |
| $s_{ij}^{xy,t}$ | Equal to 1 when branch $ij$ is connected after network reconfiguration due to a fault in branch $xy$ (or under normal operation condition if $xy$ = NO) at stage $t$. |

## I. INTRODUCTION

### A. Background

In order to improve the reliability of power supply, mesh-designed architecture is commonly adopted in the current urban distribution networks(DNs). The DN operates radially under normal conditions, and redundant lines are used for power re-routing after failures[1]-[3]. Therefore, the DN with mesh structure has higher reliability[4]. When calculating the reliability index of the mesh-designed DN, it is necessary to consider the fault isolation and load transfer after the fault which can be achieved through network reconfiguration. Otherwise, the reliability of the system may be underestimated[5]. Since different customers or lateral grids have different power supply reliability criteria, the DN expansion planning scheme is optimized to achieve the minimum investment cost constrained with specified reliability requirements. Commonly used reliability indicators include customer interruption frequency (CIF), customer interruption duration (CID), system average interruption frequency index (SAIFI) and system average interruption duration index (SAIDI)[6][7].

### B. Previous research

DN planning considering reliability has been studied. However, early researches mostly penalize the expectation of power loss in the objective function, which implicitly and approximately reflect the reliability of the DNs[8][9]. With the improvement of relevant standards, some quantitive indices are used to measure the reliability of DNs[7][10]. In order to make the planning results meet the specified reliability requirements, the reliability assessment process is required during the optimizations, which can be achieved by simulation methods or analytical methods.

Simulation-based methods often use iterative optimization-assessment procedure, that is, perform optimization steps with a posterior reliability assessment program [11]-[15]. In [11], a pool of feasible solutions with diverse expansion plans is first obtained and the reliability index of each plan is calculated to determine the optimal solution. A comprehensive planning methodology is proposed in [12] considering upgrading the conventional equipment in the DNs. The entire solution process includes the optimal DN reinforcement and the power flow solving procedure based on Gauss-Seidel iteration, which is used to evaluate the performance of the reinforcement scheme. Based on [11], the choices of customers on reliability have been considered in DN planning model by carrying out Monte Carlo-based simulation in the solution process [13]. Integrated approach for reliability planning and risk estimation in DNs proposed in [14] takes the use of backup supply or automatic/manual reconfiguration schemes into the consideration. The reliability assessment part of [14] still relies on time-sequential Monte Carlo simulation. Decision tree is used in [15] to solve multi-stage network planning problem, in which the switchgear optimization is implemented by simulation software. Tabu search is adopted to solve DN planning model considering uncertainty in [16], which requires evaluation to determine the movements for next search step. Genetic algorithm involving reliability assessment procedure to calculate fitness function is used for reliability planning stage in [17]. The simulation-based method is intuitive and easy to implement but suffers in computation time. The evaluation procedure cannot be embedded in the planning model to solve jointly while the iterative heuristic method cannot guarantee optimality.

On the other side, the analytical calculation method of reliability index has been studied in some literatures. Fault incidence matrices are formulated in [18] and the analytical representations of reliability indices with failure rate and equipment operating time is derived. A topology-variable-based reliability evaluation method is proposed in [19], which is suitable for both radial and radially-operated meshed distribution networks. In [20], a distribution network reliability assessment method considering the telecontrol and automation of switches is proposed and the reliability improvement by telecontrolled sectionalizers is verified. Islanded operation of microgrids is further considered in [21] and a general analytical expression of reliability which takes into account load shedding is derived. The probabilistic power flow is adopted to calculate the reliability index of power systems with distributed generation in [22]. Reliability index is equivalently assessed using algebraic expressions in [23], which needs less computational effort compared with simulation based method. Based on the analytical calculation model of reliability, some pioneer works have embedded analytical reliability constraints into the DN planning model. Based on fault incidence matrix proposed [24], the fault incidence matrix is applied in [25] for the joint optimization configuration of sectional switches and tie lines. Linearized models of different reliability indices are introduced in [26] and then involved in MILP model of DN planning. Network modeling formulation of reliability indices are derived in [27] and [28] to consider events such as fault isolation and load restoration in distribution system planning. The multi-level expansion planning problem of the DN is modeled in [29] as a mixed integer linear programming, which has good convergence and can be solved efficiently. However, post-fault load restoration which can improve the reliability of DNs [5] is not fully considered. For mesh DNs, planning schemes without tie switches for post-fault load restoration may lead to excessive investment.

Post-fault load restoration is considered in the DN planning model proposed in [30] but the model is solved by heuristic algorithm. Inspired by [29], a reliability-constrained distribution network planning (RcDNP) method considering post-fault load restoration for mesh DNs is proposed in [31]. The DN expansion planning problem is finally formulated as a mixed integer linear programming. However, model proposed in [31] does not take the sparse nature of the topological connection relationship of the DN into consideration. Existing studies usually model RcDNP as mixed integer linear programming. As the scale of the DN expands, the decision variables in the MILP model will explosively grow. The ultra-heavy computational burden prevents its application for real-world problems.

As an effective mean to improve solution efficiency, parallel computing has been widely applied in power system optimization, such as distributed optimal power flow [32] and distributed reactive power control [33]. However, distributed optimization algorithms adopted in the existing literature, such as ADMM [34] and ATC [35], can guarantee convergence and optimality only when the problem is convex. The MILP based DN planning model is a non-convex problem. Thus, most distributed optimization methods cannot be used directly in solving DN planning model. A parallel computing method combining branch exchange and dynamic programming for large-scale network layout optimization has been proposed in [36]. Based on [36], the simultaneous optimization of the line layout and type of conductor is further implemented in [37]. But the branch exchange algorithm adopted in network structure optimization still relies on heuristic search and is only suitable for radial DNs. In [38], a genetic algorithm based planning method is proposed considering the sparseness of the rural DN. However, heuristic method is not stable and the optimality of the solution cannot be guaranteed theoretically. Parallel accelerated solving method for large-scale mesh DN expansion planning remains to be further studied.

*C. Contributions*

Built upon [31], this paper presents a parallel computing based acceleration method for RcDNP to overcome the above difficulties. Firstly, the DN to be planned is divided into the backbone grid and lateral grids (hereafter named as sub-areas). The sub-area is a small-scale DN with relatively close topological connections. The backbone grid connects distribution transformers and various sub-regions. Secondly, according to the reliability assessment method proposed in [5], the boundary conditions satisfied by the backbone grid and sub-area variables are determined, which is calculated by the coordination layer. The RcDNP model is reformulated to the RcDNP models of the backbone grid and sub-areas. Then a parallelizable augmented Lagrangian algorithm is adopted to solve the model in parallel manner. Furthermore, the Nesterov acceleration algorithm with restart is used to improve the convergence.

The main contributions of this paper are summarized as follows:
1) A decomposition RcDNP model is proposed, in which the planning grid is decomposed into the backbone grid and several sub-area planning problems. The number of integer variables of the planning problem roughly increases linearly with the size of the planning DN, while the centralized RcDNP model in [31] increases quadratically.
2) A parallelizable augmented Lagrangian method with acceleration algorithm is developed to solve the coordination planning problem involving backbone grid and sub-areas. Numerical tests show the proposed parallel solution exhibits a linearly increasing computation time with the growing size of DNs. The optimality and convergence of the algorithm was also proved rigorously.

The remainder of this paper is arranged as follows. The decoupling planning model of backbone grid and sub-area is introduced in Section II. The parallel solution procedure with acceleration method is discussed in Section III. Numerical tests and results are demonstrated to illustrate the performance of our method in Section IV. Conclusions are drawn in Section V.

## II. MATHEMATICAL MODEL FORMULATION

*A. Distributed planning framework*

In the DN, substations are designed to supply power to multiple load concentrated areas. These load concentrated areas are connected to the substation through the backbone grid. This natural sparse structure inspires us to decompose the RcDNP problem into the backbone grid and sub-areas planning problems. It can be solved in parallel manner through coordination. The decomposition model and coordination computation framework are shown in Fig. 1.

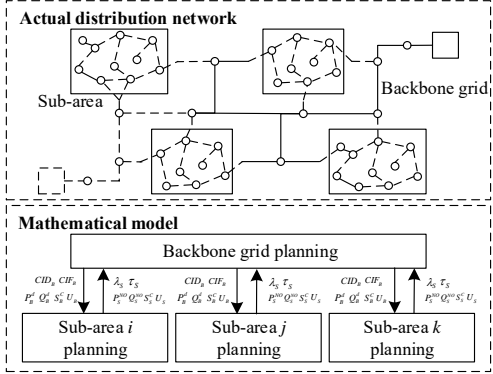

Fig. 1 Boundary shared variables between the backbone grid and sub-area

The proposed framework consists of three modules: backbone grid planning module, sub-area planning module and coordination layer. As shown in Fig 2., the sub-area is aggregated as an equivalent load node (ELN) $i'$ in the backbone grid planning problem. For the sub-area planning problem, the backbone grid is represented by an equivalent distribution source (EDS) $i''$ connected a series equivalent outlet branch (EOB) $i''j''$ with a certain probability of failure.

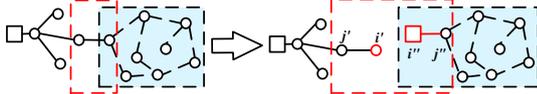

Fig. 2 Schematic diagram of equivalent decoupled models of backbone grid and sub-area for RcDNP

The consistency conditions of the boundary variables include two aspects: the consistency of the power flow and the consistency of the reliability index.

The consistency of the power flow means that the equivalent load of the backbone grid should be equal to the power flow of corresponding EOB in the sub-area. Besides, corresponding branches in the backbone grid and sub-area share the same capacity while corresponding nodes in the backbone grid and sub-area share the same voltage.

$$P_{i'}^{d,t} = P_{i''j''}^{NO,t}, i' \in \Psi_D^e, i'' \in \Psi_S^e, j'' \in \Psi_{i''} \quad (1)$$

$$Q_{i'}^{d,t} = Q_{i''j''}^{NO,t}, i' \in \Psi_D^e, i'' \in \Psi_S^e, j'' \in \Psi_{i''} \quad (2)$$

$$S_{i'j'}^{C,t} = S_{i''j''}^{C,t}, i' \in \Psi_D^e, i'' \in \Psi_S^e, j' \in \Psi_{i'}, j'' \in \Psi_{i''} \quad (3)$$

$$U_{i'}^{NO,t} = U_{i''}^{ss,t}, i' \in \Psi_D^e, i'' \in \Psi_S^e \quad (4)$$

Here $\Psi_D^e$ is the set of equivalent load node in the backbone grid and $\Psi_S^e$ is the set of equivalent source node in the sub-area.

The consistency of the reliability index means that the parameter of the EOB in the sub-area and the reliability index of the ELN of backbone grid should meet the following constraints.

$$\lambda_{i''j''}^t = CIF_{i'}^{e,t}, i' \in \Psi_D^e, i'' \in \Psi_S^e, j'' \in \Psi_{i''} \quad (5)$$

$$\tau_{i''j''}^t = CID_{i'}^{e,t} / CIF_{i'}^{e,t}, i' \in \Psi_D^e, i'' \in \Psi_S^e, j'' \in \Psi_{i''} \quad (6)$$

Here $CID_{i'}^{e,t}$ and $CIF_{i'}^{e,t}$ are the temporarily estimated customer interruption duration and customer interruption frequency of the ELN $i'$ in the backbone grid planning module. The EDS $i''$ is assumed completely reliable, the influence of the backbone grid' failures on the sub-area is reflected in the EOB.

*B. Objective function*

The objective function for the network expansion planning model is the total investment which consists of the investment cost, the maintenance cost and the reliability cost.

$$\min f(\mathbf{x}) = \sum_{t=1}^{T} \left[ \delta_t^I c_t^I + \delta_t^O (c_t^M + \omega EENS^t) \right] \quad (7)$$

The investment cost $c_t^I$ and the maintenance cost $c_t^M$ for stage $t$ are defined as followed.

$$c_t^I = \sum_{ij \in \Psi_B} \sum_{a \in \Lambda^C} C_C^{a,I} l_{ij}^{a,t} + \sum_{Tr \in \Psi_T} \sum_{a \in \Lambda^T} C_T^{a,I} m_{Tr}^{a,t} + \sum_{s \in \Psi_{SS}} C_S^{s,I} n_s^t \quad (8)$$

$$c_t^M = \boldsymbol{e}_t^T \sum_{ij \in \Psi_B} \sum_{a \in \Lambda^C} C_C^{a,M} (\sum_{\tau=1}^t l_{ij}^{a,\tau} \boldsymbol{g}_C^{a,\tau} + l_{ij}^{a,0} \boldsymbol{g}_C^{a,0})$$

$$+ \boldsymbol{e}_t^T \sum_{Tr \in \Psi_T} \sum_{a \in \Lambda^T} C_T^{a,M} (\sum_{\tau=1}^t m_{Tr}^{a,\tau} \boldsymbol{g}_T^{a,\tau} + m_{Tr}^{a,0} \boldsymbol{g}_T^{a,0}) \quad (9)$$

$$+ \sum_{s \in \Psi_{SS}} C_S^{s,M} (\sum_{\tau=1}^t n_s^t + n_s^0)$$

The present value factor for investment and maintenance costs at stage $t$ are

$$\delta_t^I = \frac{1}{(1+I)^{\tau(t)}}, \delta_t^O = \sum_{\tau=\tau(t)}^{\tau(t+1)-1} \frac{1}{(1+I)^\tau} \quad (10)$$

Where $C_C^{a,I}$ and $C_C^{a,M}$ are the investment cost and maintenance cost for alternative conductor $a \in \Lambda^C$. $C_T^{a,I}$ and $C_T^{a,M}$ the investment cost and maintenance cost for alternative transformer $a \in \Lambda^T$. $C_S^{s,I}$ and $C_S^{s,M}$ is the investment cost and maintenance cost for substation at node $s$. Binaries $l_{ij}^{a,t}$, $m_{Tr}^{a,t}$ and $n_s^t$ represent that whether to install alternative conductor $a$ at branch $ij$, whether to install alternative transformer $a$ at candidate position $Tr$ and whether to build new substation at node $s$ in stage $t$. $\tau(t)$ is the number of year up to stage $t$. $\boldsymbol{g}_C^{a,\tau}$, $\boldsymbol{g}_C^{a,\tau}$, $\boldsymbol{g}_T^{a,\tau}$ and $\boldsymbol{g}_T^{a,0}$ are aging vectors for conductors and transformers, which is shown as followed.

$$\boldsymbol{g}_C^{a,\tau} = [0 \cdots 0 \underset{\tau}{1} \cdots \underset{\tau+T_C^a-1}{1} 0 \cdots 0]^T, \boldsymbol{g}_C^{a,0} = [1 \cdots \underset{T_C^{a,0}}{1} 0 \cdots 0]^T$$

$$\boldsymbol{g}_T^{a,\tau} = [0 \cdots 0 \underset{\tau}{1} \cdots \underset{\tau+T_T^a-1}{1} 0 \cdots 0]^T, \boldsymbol{g}_C^{a,0} = [1 \cdots \underset{T_T^{a,0}}{1} 0 \cdots 0]^T \quad (11)$$

Where $T_C^a$ is the number of stage in the lifespan of the alternative conductor $a \in \Lambda^C$, and $T_C^{a,0}$ is the number of stage in the remaining lifespan of the existing alternative conductor $a \in \Lambda^C$. $T_T^a$ is the number of stage in the lifespan of the alternative transformer $a \in \Lambda^T$, and $T_T^{a,0}$ is the number of stage in the remaining lifespan of the existing alternative transformer $a \in \Lambda^T$.

*C. Constraints*

The constraints in the model include:

*1) Operational constraints*

Operational constraints are classified into normal conditions and fault conditions.

$$\sum_{j \in \Psi_i} P_{ij}^{xy,t} + P_i^{xy,t} + P_{i,g}^{xy,t} = 0, \forall i \in \Psi_N \quad (12)$$

$$\sum_{j \in \Psi_i} Q_{ij}^{xy,t} + Q_i^{xy,t} + Q_{i,g}^{xy,t} = 0, \forall i \in \Psi_N \quad (13)$$

$$-M(1-s_{ij}^{xy,t}) \leq U_j^{xy,t} - U_i^{xy,t} + 2(r_{ij}^t P_{ij}^{xy,t} + x_{ij}^t Q_{ij}^{xy,t}), \forall ij \in \Psi_B \quad (14)$$

$$U_j^{xy,t} - U_i^{xy,t} + 2(r_{ij}^t P_{ij}^{xy,t} + x_{ij}^t Q_{ij}^{xy,t}) \leq M(1-s_{ij}^{xy,t}), \forall ij \in \Psi_B \quad (15)$$

$$\underline{U} \leq U_i^{xy,t} \leq \overline{U}, \forall i \in \Psi_N \quad (16)$$

$$U_i^{xy,t} = U_i^{ss,t}, \forall i \in \Psi_{SS} \quad (17)$$

$$P_{Tr}^{xy,t} = P_{ij}^{xy,t}, Tr \in \Psi_T, ij \in \Psi_{Tr}^{out} \quad (18)$$

$$Q_{Tr}^{xy,t} = Q_{ij}^{xy,t}, Tr \in \Psi_T, ij \in \Psi_{Tr}^{out} \quad (19)$$

$$\begin{cases} -Ms_{ij}^{xy,t} \leq P_{ij}^{xy,t} \leq Ms_{ij}^{xy,t} \\ -Ms_{ij}^{xy,t} \leq Q_{ij}^{xy,t} \leq Ms_{ij}^{xy,t} \end{cases}, \forall ij \in \Psi_B \quad (20)$$

$$\begin{cases} -S_{ij}^{C,t} \leq P_{ij}^{xy,t} \leq S_{ij}^{C,t} \\ -S_{ij}^{C,t} \leq Q_{ij}^{xy,t} \leq S_{ij}^{C,t} \end{cases}, \forall ij \in \Psi_B \quad (21)$$

$$-\sqrt{2}S_{ij}^{C,t} \leq P_{ij}^{xy,t} \pm Q_{ij}^{xy,t} \leq \sqrt{2}S_{ij}^{C,t}, \forall ij \in \Psi_B \quad (22)$$

$$\begin{cases} P_{Tr}^{xy,t} \leq S_{Tr}^{C,t} \\ Q_{Tr}^{xy,t} \leq S_{Tr}^{C,t} \end{cases}, \forall Tr \in \Psi_T \quad (23)$$

$$-\sqrt{2}S_{Tr}^{C,t} \leq P_{Tr}^{xy,t} \pm Q_{Tr}^{xy,t} \leq \sqrt{2}S_{Tr}^{C,t}, \forall Tr \in \Psi_T \quad (24)$$

$$s_{ij}^{xy,t} \leq l_{ij}^t, \forall ij \in \Psi_B \quad (25)$$

Here $xy \in \Psi_B \cup \{NO\}$ represents branch fault set and normal operation. Constraints (12)(13) indicate the balance between the power injection and outflow. Constraints (14)(15) are the linearized power flow constraints. Constraints (16)(17) are the nodal voltage constraints. Constraints (18)(19) indicate that the power flow of transformer and its outlet branch are consistent, that is the power flow of the transformer is equal to the power flow of its outlet branch. (20) is constraint of the power flow over branch related to branch status. (21)(22) and (23)(24) restrict the power flow of branch and transformer from exceeding maximum transmission apparent power to ensure the safe operation. Constraint (25) indicates the connecting status equal to zero when there is no branch constructed.

2) *Fault indicator variable constraints*

$$s_{xy}^{xy,t} = 0 \quad (26)$$

$$h_i^{f,t} + h_{xy}^{f,t} - 1 \leq p_i^{xy,t}, \forall f \in \Psi_F, \forall i \in \Psi_N \quad (27)$$

$$p_i^{xy,t} \geq q_i^{xy,t}, \forall i \in \Psi_N \quad (28)$$

$$P_i^{xy,t} = P_i^t (1 - q_i^{xy,t}), \forall i \in \Psi_N \quad (29)$$

$$Q_i^{xy,t} = Q_i^t (1 - q_i^{xy,t}), \forall i \in \Psi_N \quad (30)$$

$$\sum_{ij} s_{ij}^{xy,t} = \sum_i (1 - q_i^{xy,t}), ij \in \Psi_B, i \in \Psi_N \text{ and } i \notin \Psi_{SS} \quad (31)$$

$$-M(1-s_{ij}^{NO,t}) + h_i^{f,t} \leq h_{ij}^{f,t} \leq h_i^{f,t} + M(1-s_{ij}^{NO,t}), \forall ij \in \Psi_B, \forall f \in \Psi_F \quad (32)$$

$$-M(1-s_{ij}^{NO,t}) + h_j^{f,t} \leq h_{ij}^{f,t} \leq h_j^{f,t} + M(1-s_{ij}^{NO,t}), \forall ij \in \Psi_B, \forall f \in \Psi_F \quad (33)$$

$$h_{ij}^{f,t} = s_{ij}^{NO,t}, \text{if line } ij \text{ is outlet branch of feeder } f \quad (34)$$

$$h_{ij}^{f,t} \leq s_{ij}^{NO,t}, \forall ij \in \Psi_B, \forall f \in \Psi_F \quad (35)$$

$$\sum_f h_i^{f,t} \leq 1, \forall i \in \Psi_N \quad (36)$$

$$\sum_f h_{ij}^{f,t} \leq 1, \forall ij \in \Psi_B \quad (37)$$

$$\sum_{ij} s_{ij}^{NO,t} = \sum_f \sum_i h_i^{f,t}, ij \in \Psi_B, f \in \Psi_F, i \in \Psi_N \text{ and } i \notin \Psi_{SS} \quad (38)$$

Here $xy \in \Psi_B$ represent failure scenario. Constraint (26) indicates that branch $xy$ is outage and isolated in the scenario where branch $xy$ fails. Constraints (27) determines that the affected nodes by outage of branch $xy$ must share the same feeder affiliation variable with the one of branch $xy$. Constrains (28) indicates nodes that are not affected by the fault cannot loss power supply due to network reconfiguration. Constraints (29)-(30) indicates that if the node can restore power supply after the post-fault network reconfiguration, its load level returns to normal state, otherwise it remains in the outage state. (32)-(35) are constraints of the feeder affiliation variables related to network topology in normal state, where constrains (32)-(33) show that feeder affiliation variables can be propagated if branch $ij$ is connected under normal operating conditions. (31) and (36)-(38) are radial operation constraints under normal and fault conditions.

Regarding the problem of backbone grid planning, the constraints (29)(30) needs to be replaced with the following constraints.

$$P_i^t = \begin{cases} P_i^{NO,t}, \forall i \in \Psi_N \text{ and } i \notin \Psi_D^e \\ P_i^{d,t}, \forall i \in \Psi_N \text{ and } i \in \Psi_D^e \end{cases} \quad (39)$$

$$Q_i^t = \begin{cases} Q_i^{NO,t}, \forall i \in \Psi_N \text{ and } i \notin \Psi_D^e \\ Q_i^{d,t}, \forall i \in \Psi_N \text{ and } i \in \Psi_D^e \end{cases} \quad (40)$$

Here $P_i^{d,t}$ and $Q_i^{d,t}$ are the equivalent load of sub-area at node $i$ at stage $t$.

3) *Equipment selection constraints*

$$l_{ij}^t = \mathbf{e}_t^T \sum_{a \in \Lambda^C} (\sum_{\tau=1}^t l_{ij}^{a,\tau} \mathbf{g}_C^{a,\tau} + l_{ij}^{a,0} \mathbf{g}_C^{a,0}), \forall ij \in \Psi_B \quad (41)$$

$$S_{ij}^{C,t} = \mathbf{e}_t^T \sum_{a \in \Lambda^C} S_C^a (\sum_{\tau=1}^t l_{ij}^{a,\tau} \mathbf{g}_C^{a,\tau} + l_{ij}^{a,0} \mathbf{g}_C^{a,0}), \forall ij \in \Psi_B \quad (42)$$

$$r_{ij}^t = d_{ij} \mathbf{e}_t^T \sum_{a \in \Lambda^C} r_C^a (\sum_{\tau=1}^t l_{ij}^{a,\tau} \mathbf{g}_C^{a,\tau} + l_{ij}^{a,0} \mathbf{g}_C^{a,0}), \forall ij \in \Psi_B \quad (43)$$

$$x_{ij}^t = d_{ij} \mathbf{e}_t^T \sum_{a \in \Lambda^C} x_C^a (\sum_{\tau=1}^t l_{ij}^{a,\tau} \mathbf{g}_C^{a,\tau} + l_{ij}^{a,0} \mathbf{g}_C^{a,0}), \forall ij \in \Psi_B \quad (44)$$

$$\lambda_{ij}^t = d_{ij} \mathbf{e}_t^T \sum_{a \in \Lambda^C} \lambda_C^a (\sum_{\tau=1}^t l_{ij}^{a,\tau} \mathbf{g}_C^{a,\tau} + l_{ij}^{a,0} \mathbf{g}_C^{a,0}), \forall ij \in \Psi_B \quad (45)$$

$$m_{Tr}^t = \mathbf{e}_t^T \sum_{a \in \Lambda^T} (\sum_{\tau=1}^t m_{Tr}^{a,\tau} \mathbf{g}_T^{a,\tau} + m_{Tr}^{a,0} \mathbf{g}_T^{a,0}), \forall Tr \in \Psi_T \quad (46)$$

$$S_{Tr}^{C,t} = \mathbf{e}_t^T \sum_{a \in \Lambda^T} S_{Tr}^a \left( \sum_{\tau=1}^t m_{Tr}^{a,\tau} \mathbf{g}_T^{a,\tau} + m_{Tr}^{a,0} \mathbf{g}_T^{a,0} \right), \forall Tr \in \Psi_T \quad (47)$$

$$\sum_{\tau=1}^t n_s^\tau + n_s^0 \leq 1, \forall s \in \Psi_{SS} \quad (48)$$

$$\sum_{\tau=1}^t \sum_{Tr \in \Psi_T} m_{Tr}^\tau \leq N_s (\sum_{\tau=1}^t n_s^\tau + n_s^0), \forall s \in \Psi_{SS} \quad (49)$$

Here constraints (41)-(48) indicates available equipment must already exist or be constructed at the planning stage. Logic constraints between the installation of transformers and the existence of substation are demonstrated as constraints (49).

4) *Reliability index calculation*

$$CIF_i^t = \sum_{xy \in \Psi_B} \lambda_{xy}^t p_i^{xy,t}, \forall i \in \Psi_N \quad (50)$$

$$CID_i^t = \sum_{xy \in \Psi_B} \lambda_{xy}^t \tau_{xy}^{SW} p_i^{xy,t} + \sum_{xy \in \Psi_B} \lambda_{xy}^t \left( \tau_{xy}^{RP} - \tau_{xy}^{SW} \right) q_i^{xy,t}, \forall i \in \Psi_N \quad (51)$$

$$SAIDI^t = \sum_{i \in \Psi_N} NC_i^t CID_i^t \Big/ \sum_{i \in \Psi_N} NC_i^t \quad (52)$$

$$SAIDI^t \leq \varepsilon_{SAIDI}^t \quad (53)$$

Constraints (50)-(52) is the common used reliability indices and constraint (53) is the reliability constraint expressed by SAIDI. For the sub-area planning problem, since the failure rate and repair time of the EOB are also variables, the constraint (51) should be rewritten as

$$CID_i^t = \sum_{\substack{x \notin \Psi_S^e \\ y \in \Psi_x}} \lambda_{xy}^t \tau_{xy}^{SW} p_i^{xy,t} + \sum_{\substack{x \notin \Psi_S^e \\ y \in \Psi_x}} \lambda_{xy}^t \left(\tau_{xy}^{RP} - \tau_{xy}^{SW}\right) q_i^{xy,t} + \sum_{\substack{x \in \Psi_S^e \\ y \in \Psi_x}} \lambda_{xy}^t \tau_{xy}^t p_i^{xy,t}, \forall i \in \Psi_N \quad (54)$$

Here $xy \in \Psi_B$ represent branch fault set. $\Psi_S^e$ is the set of equivalent source node in the sub-area. The second row of constraint (54) is related to the equivalent outlet branch of the backbone grid, the failure rate $\lambda_{xy}^t$ and repair time $\tau_{xy}^t$ of which are boundary variables. Then big M method is applied to deal with the product terms $\lambda_{xy}^t \tau_{xy}^t p_i^{xy,t}$ in the second line of constraint (54).

$$CID_i^t = \sum_{\substack{x \notin \Psi_S^e \\ y \in \Psi_x}} \lambda_{xy}^t \tau_{xy}^{SW} p_i^{xy,t} + \sum_{\substack{x \notin \Psi_S^e \\ y \in \Psi_x}} \lambda_{xy}^t \left(\tau_{xy}^{RP} - \tau_{xy}^{SW}\right) q_i^{xy,t} + \sum_{\substack{x \in \Psi_S^e \\ y \in \Psi_x}} n_i^{xy,t}, \forall i \in \Psi_N \quad (55)$$

$$\begin{cases} n_i^{xy,t} \geq u_{xy}^t - (1 - p_i^{xy,t})M \\ n_i^{xy,t} \geq -p_i^{xy,t}M \\ n_i^{xy,t} \leq u_{xy}^t + (1 - p_i^{xy,t})M \\ n_i^{xy,t} \leq p_i^{xy,t}M \end{cases}, x \in \Psi_S^e, y \in \Psi_x \quad (56)$$

Here $u_{xy}^t$ is the substitution variable for the product of the failure rate $\lambda_{xy}^t$ and repair time $\tau_{xy}^t$ of the EOB.

### III. SOLUTION PROCEDURE

This section introduces a parallelizable augmented Lagrangian algorithm [39], which is applicable for split-variable reformulation of mixed-integer optimization problems, is adopted to solve the coordination planning problem of backbone grid and sub-areas. The backbone grid and sub-areas achieve independent planning by solving sub-problems in parallel. The global coordination is achieved by the iteration between the coordination layer and sub-problems.

*A. Decomposable RcDNP model*

The algorithm presented in reference [39] is adapted to solve problems with following form:

$$\min_{x_i, z_i, \forall i} \left\{ \sum_{i=1}^n f_i(x_i) : Q_i x_i = z_i, x_i \in X_i, \forall i, (z_1^T, ..., z_n^T)^T \in Z \right\} \quad (57)$$

The model presented in the previous reference is a centralized model, which need to be reorganized to adapt to the decomposition calculation method. Combining the model of backbone grid and sub-area, we can obtain the decomposable RcDNP model of the entire system:

$$\min \; f(x_b, x_s) = f_b(x_b) + \sum_{n=1}^{n_{sub}} f_{s,n}(x_{s,n})$$

s.t. (12)-(28),(31)-(52), *for backbone grid*

(12)-(38),(41)-(50),(52)-(53),(55)-(56), *for sub-area*

(1)-(6)

With variable substitution in constraint (56), the constraints (5) and (6) can be written as the following form.

$$\lambda_{i'j''}^t = CIF_{i'}^{e,t}, i' \in \Psi_D^e, i'' \in \Psi_S^e, j'' \in \Psi_{i''} \quad (58)$$

$$u_{i'j''}^t = CID_{i'}^{e,t}, i' \in \Psi_D^e, i'' \in \Psi_S^e, j'' \in \Psi_{i''} \quad (59)$$

To further organize the model into a form suitable for augmented Lagrangian methods, the decomposable RcDNP model can be written as the compact matrix form:

$$\min \; f(x_b, x_s) = f_b(x_b) + \sum_{n=1}^{n_{sub}} f_{s,n}(x_{s,n})$$

$$s.t. \; x_b \in X_b, x_{s,n} \in X_{s,n} \quad (60)$$

$$Q_b x_b = Q_{s,n} x_{s,n}, n = 1, ..., n_{sub}$$

Here $X_b$ is a non-convex feasible region of backbone grid planning model constructed by the mixed-integer linear constraints (12)-(28) and (31)-(52). $X_{s,n}$ is a non-convex feasible region of sub-area planning model constructed by the mixed-integer linear constraints (12)-(38),(41)-(50),(52)-(53) and (55)-(56). The second row of the constraints represents coordination constraints between backbone grid and sub-areas.

In order to make the structure of the model clearer, we further define $z_b$ and $z_{s,n}$ as

$$Q_b x_b = z_b = \left(CIF_{i'}^{e,t}, CID_{i'}^{e,t}, P_{i'}^{d,t}, Q_{i'}^{d,t}, S_{i'j'}^{C,t}, U_{i'}^{NO,t}\right)^T, \forall i' \in \Psi_D^e, j' \in \Psi_{i'} \quad (61)$$

$$Q_{s,n} x_{s,n} = z_{s,n} = \left(\lambda_{i''j''}^t, u_{i''j''}^t, P_{i''j''}^{NO,t}, Q_{i''j''}^{NO,t}, S_{i''j''}^{C,t}, U_{i''}^{ss,t}\right)^T, \quad (62)$$

$$i'' \in \Psi_{S,n}^e, j'' \in \Psi_{i''}, n = 1, 2, ..., n_{sub}$$

Finally, set Z is defined to describe the coupling relationship between the coordination variables of different models, which is restricted by coordination constraints (1)-(4) and (58)-(59).

$$Z = \left\{ (z_b^T, z_{s,1}^T, ..., z_{s,n_{sub}}^T)^T \mid (1),(2),(3),(4),(58),(59) \right\} \quad (63)$$

The coupling vectors of backbone grid $z_b$ and sub-area $z_{s,n}$ are confined to regions constructed by coordination constraints. Thus, a decomposable RcDNP model is derived in the form of problem (57).

$$\min_{x,z} \; f(x_b, x_s) = f_b(x_b) + \sum_{n=1}^{n_{sub}} f_{s,n}(x_{s,n})$$

$$s.t. \; Q_b x_b = z_b, x_b \in X_b, \quad (64)$$

$$Q_{s,n} x_{s,n} = z_{s,n}, x_{s,n} \in X_{s,n},$$

$$(z_b^T, z_{s,1}^T, ..., z_{s,n_{sub}}^T)^T \in Z$$

So the augmented Lagrangian method can be adopted to solve problem (64). The detailed solution steps are described as followed.

**Step 1. Initialization**

Define the augmented Lagrangian function as

$$L_b(x_b, w_b, z_b) = f(x_b) + w_b^T Q_b x_b + \frac{\rho}{2} \|Q_b x_b - z_b\|^2 \quad (65)$$

$$L_{s,n}(x_{s,n}, w_{s,n}, z_{s,n}) = f(x_{s,n}) + w_{s,n}^T Q_{s,n} x_{s,n} + \frac{\rho}{2} \|Q_{s,n} x_{s,n} - z_{s,n}\|^2 \quad (66)$$

Here $w_b$ and $w_{s,n}$ are the Lagrangian multiplier vectors, $\rho$ is a penalty value. Initialize parameters $\underline{\phi}_b = \underline{\phi}_{s,n} = -\infty$, $\rho > 0$, $\varepsilon > 0$, $k = 0$, $T = 0$, $\gamma \in (0,1)$.

**Step 2. Solve the initial value of the sub-problem**

Solve the problem of the backbone grid (P1).

Obj. $\quad f_b(\boldsymbol{x}_b) = \sum_{t=1}^{T}\left[\delta_t^I c_{t,b}^I + \delta_t^O(c_{t,b}^M + \omega EENS_b^t)\right]$

s.t.
$$\underline{P_{i'}^d} \leq P_{i'}^{d,t} \leq \overline{P_{i'}^d}, i' \in \Psi_D^e \quad (67)$$
$$\underline{Q_{i'}^d} \leq Q_{i'}^{d,t} \leq \overline{Q_{i'}^d}, i' \in \Psi_D^e \quad (68)$$
$$\underline{CIF_{i'}^e} \leq CIF_{i'}^{e,t} \leq \overline{CIF_{i'}^e}, i' \in \Psi_D^e \quad (69)$$
$$\underline{CID_{i'}^e} \leq CID_{i'}^{e,t} \leq \overline{CID_{i'}^e}, i' \in \Psi_D^e \quad (70)$$
(12)-(28),(31)-(52)

Solve the planning model of each sub-area (P2).

Obj. $\quad f_{s,n}(\boldsymbol{x}_{s,n}) = \sum_{t=1}^{T}\left[\delta_t^I c_{t,s,n}^I + \delta_t^O(c_{t,s,n}^M + \omega EENS_{s,n}^t)\right]$

s.t.
$$\underline{P_{i''j''}^{NO}} \leq P_{i''j''}^{NO,t} \leq \overline{P_{i''j''}^{NO}}, i'' \in \Psi_{S,n}^e, j'' \in \Psi_{i''} \quad (71)$$
$$\underline{Q_{i''j''}^{NO}} \leq Q_{i''j''}^{NO,t} \leq \overline{Q_{i''j''}^{NO}}, i'' \in \Psi_{S,n}^e, j'' \in \Psi_{i''} \quad (72)$$
$$\underline{\lambda_{i''j''}} \leq \lambda_{i''j''}^t \leq \overline{\lambda_{i''j''}}, i'' \in \Psi_{S,n}^e, j'' \in \Psi_{i''} \quad (73)$$
$$\underline{u_{i''j''}} \leq u_{i''j''}^t \leq \overline{u_{i''j''}}, i'' \in \Psi_{S,n}^e, j'' \in \Psi_{i''} \quad (74)$$
(12)-(38),(41)-(50),(52)-(53),(55)-(56)

Construct the convex hull of the feasible region $\boldsymbol{D}_b = \{\boldsymbol{x}_b^0\}$ and $\boldsymbol{D}_{s,n} = \{\boldsymbol{x}_{s,n}^0\}$ using the linear combination of above solution. Here $\boldsymbol{x}_b^0$ is the solution of the backbone grid planning problem and $\boldsymbol{x}_{s,n}^0$ is the solution of $n$th sub-area planning problem.

The coordination layer calculates coordination variables by solving the following optimization problem (P3).

$$z^0 = \arg\min_{z_b, z_{s,n}}\{\|Q_b \boldsymbol{x}_b^0 - z_b\|^2 + \sum_{n=1}^{n_{sub}}\|Q_{s,n}\boldsymbol{x}_{s,n}^0 - z_{s,n}\|^2 : z_b, z_{s,n} \in Z\}$$

Here $z^0 = (z_b^0, z_{s,1}^0, ..., z_{s,n_{sub}}^0)$ is the set of coordination variables for the backbone grid and all sub-area planning problems.

**Step 3.** Update: $k = k + 1$. T=1.

**Step 4. Nonlinear Gauss–Seidel iteration**

Update variables: $\boldsymbol{x}_b^k \leftarrow \boldsymbol{x}_b^{k-1}$, $\boldsymbol{x}_{s,n}^k \leftarrow \boldsymbol{x}_{s,n}^{k-1}$, $z^k \leftarrow z^{k-1}$, $w_b^k \leftarrow w_b^{k-1}$, $w_{s,n}^k \leftarrow w_{s,n}^{k-1}$, $\underline{\phi}^k \leftarrow \underline{\phi}^{k-1}$.

Solve the optimization model for the backbone grid (P4):
$$\boldsymbol{x}_b^k \leftarrow \arg\min_{\boldsymbol{x}_b}\{L_b(\boldsymbol{x}_b, w_b^k, z_b^k) : \boldsymbol{x}_b \in \boldsymbol{D}_b\}$$

Solve the optimization model of each sub-area (P5):
$$\boldsymbol{x}_{s,n}^k \leftarrow \arg\min_{\boldsymbol{x}_{s,n}}\{L_{s,n}(\boldsymbol{x}_{s,n}, w_{s,n}^k, z_{s,n}^k) : \boldsymbol{x}_{s,n} \in \boldsymbol{D}_{s,n}\}$$

Solve the model P3 to update the coordination variables:
$$z^k = \arg\min_{z_b, z_{s,n}}\{\|Q_b \boldsymbol{x}_b^k - z_b\|^2 + \sum_{n=1}^{n_{sub}}\|Q_{s,n}\boldsymbol{x}_{s,n}^k - z_{s,n}\|^2 : z_b, z_{s,n} \in Z\}$$

**Step 5.** T=T+1. If $T \leq T_{\max}$, return to Step 4. Otherwise, perform the following steps.

$$\tilde{\phi}_b \leftarrow \underline{\phi}_b(w_b^k + \rho(Q_b \boldsymbol{x}_b^k - z_b^k)), \tilde{\phi}_{s,n} \leftarrow \underline{\phi}_{s,n}(w_{s,n}^k + \rho(Q_{s,n}\boldsymbol{x}_{s,n}^k - z_{s,n}^k))$$

Here $\underline{\phi}_b(w_b)$ and $\underline{\phi}_{s,n}(w_{s,n})$ are defined as the dual function of the original problems:

$$\underline{\phi}_b(w_b) = \min_{\boldsymbol{x}_b}\{f_b(\boldsymbol{x}_b) + w_b^T Q_b \boldsymbol{x}_b : \boldsymbol{x}_b \in X_b\} \quad (75)$$

$$\underline{\phi}_{s,n}(w_{s,n}) = \min_{\boldsymbol{x}_{s,n}}\{f_{s,n}(\boldsymbol{x}_{s,n}) + w_{s,n}^T Q_{s,n}\boldsymbol{x}_{s,n} : \boldsymbol{x}_{s,n} \in X_{s,n}\} \quad (76)$$

Solve the optimization model of the backbone grid (P6):
$$\hat{\boldsymbol{x}}_b \leftarrow \arg\min_{\boldsymbol{x}_b}\{\underline{\phi}_b(w_b^k + \rho(Q_b \boldsymbol{x}_b^k - z_b^k)\}$$

Solve the optimization model of each sub-area (P7):
$$\hat{\boldsymbol{x}}_{s,n} \leftarrow \arg\min_{\boldsymbol{x}_{s,n}}\{\underline{\phi}_{s,n}(w_{s,n}^k + \rho(Q_{s,n}\boldsymbol{x}_{s,n}^k - z_{s,n}^k)\}$$

Add the vertex to the convex hull of feasible region:
$$\boldsymbol{D}_b \leftarrow conv(\boldsymbol{D}_b \cup \{\hat{\boldsymbol{x}}_b\}), \boldsymbol{D}_{s,j} \leftarrow conv(\boldsymbol{D}_{s,n} \cup \{\hat{\boldsymbol{x}}_{s,n}\})$$

$$\varepsilon_b^k \leftarrow \hat{\phi}_b(w_b^k, \boldsymbol{x}_b^k, z_b^k) - \underline{\phi}_b^k, \varepsilon_{s,n}^k \leftarrow \hat{\phi}_{s,n}(w_{s,n}^k, \boldsymbol{x}_{s,n}^k, z_{s,n}^k) - \underline{\phi}_{s,n}^k$$

$$\Delta \phi_b^k \leftarrow \tilde{\phi}_b - \underline{\phi}_b^k, \Delta \phi_{s,n}^k \leftarrow \tilde{\phi}_{s,n} - \underline{\phi}_{s,n}^k$$

Here $\hat{\phi}_b(w_b, \boldsymbol{x}_b, z_b)$ and $\hat{\phi}_{s,n}(w_{s,n}, \boldsymbol{x}_{s,n}, z_{s,n})$ are cutting plane functions used to approximate the dual function (75) and (76), which are defined as followed.

$$\hat{\phi}_b(w_b, \boldsymbol{x}_b, z_b) = L_b(\boldsymbol{x}_b, w_b, z_b) + \frac{\rho}{2}\|Q_b \boldsymbol{x}_b - z_b\|^2 \quad (77)$$

$$\hat{\phi}_{s,n}(w_{s,n}, \boldsymbol{x}_{s,n}, z_{s,n}) = L_{s,n}(\boldsymbol{x}_{s,n}, w_{s,n}, z_{s,n}) + \frac{\rho}{2}\|Q_{s,n}\boldsymbol{x}_{s,n} - z_{s,n}\|^2 \quad (78)$$

The algorithm converges if the gap $\varepsilon_b^k$ and $\varepsilon_{s,n}^k$ is small enough.

**Step 6. Convergence criterion**

If $\varepsilon_b^k + \sum_{n=1}^{n_{sub}} \varepsilon_{s,n}^k \leq \varepsilon$, algorithm converges. Otherwise, perform the following steps.

Calculate $\eta^k = \Delta \phi_b^k / \varepsilon_b^k + \sum_{n=1}^{n_{sub}} \Delta \phi_{s,n}^k / \varepsilon_{s,n}^k$. If $\eta^k \geq \gamma$, update $w_b^k$ and $w_{s,n}^k$:

$$w_b^k \leftarrow w_b^k + \rho(Q_b \boldsymbol{x}_b^k - z_b^k), w_{s,n}^k \leftarrow w_{s,n}^k + \rho(Q_{s,n}\boldsymbol{x}_{s,n}^k - z_{s,n}^k)$$

$$\underline{\phi}_b^k \leftarrow \tilde{\phi}_b, \underline{\phi}_{s,n}^k \leftarrow \tilde{\phi}_{s,n}$$

**Step 7.** Update $\rho$

$$\frac{1}{\rho} \leftarrow \min\left\{\max\left\{\frac{2}{\rho}(1-\eta^k), \frac{1}{10\rho}, \frac{1}{\rho_{\max}}\right\}, \frac{10}{\rho}, \frac{1}{\rho_{\min}}\right\}$$

If $k > k_{\max}$, algorithm stop. Otherwise, return to Step 3.

**B. Acceleration method**

In order to speed up the solution, we apply the Nesterov acceleration algorithm in the solution procedure. To enhance its stability, a restart rule is adopted as following.

**Algo 1.** Nesterov acceleration algorithm with restart.

**Initialization:**

1: **For** $k=1,2,3,...,k_{\max}$ do

2: $\quad \boldsymbol{x}_b^k = \arg\min_{\boldsymbol{x}_b} f_b(\boldsymbol{x}_b) + \hat{w}_b^k Q_b \boldsymbol{x}_b + \rho\|Q_b \boldsymbol{x}_b - z_b\|^2 / 2$

3:     $w_b^k = \hat{w}_b^k + \rho(Q_b x_b^k - z_b^k)$

4: **For** $n=1,2,3,\ldots,n_{sub}$ **do**

5:     $x_{s,n}^k = \arg\min_{x_{s,n}} f_{s,n}(x_{s,n}) + \hat{w}_{s,n}^k Q_{s,n} x_{s,n} + \rho \| Q_{s,n} x_{s,n} - z_{s,n} \|^2 / 2$

6:     $w_{s,n}^k = \hat{w}_{s,n}^k + \rho(Q_{s,n} x_{s,n}^k - z_{s,n}^k)$

7: **end for**

8: $w^k = (w_b^k, w_{s,1}^k, \ldots, w_{s,n_{sub}}^k), \hat{w}^k = (\hat{w}_b^k, \hat{w}_{s,1}^k, \ldots, \hat{w}_{s,n_{sub}}^k)$

9: $c_k = \rho^{-1} \| w_k - \hat{w}_k \|^2$

10: **if** $c_k < \delta c_{k-1}$ **then**

11:     $\alpha_{k+1} = (1 + \sqrt{1 + 4\alpha_k^2})/2, \hat{w}_{k+1} = w_k + (\alpha_k - 1)(w_k - w_{k-1})/\alpha_{k+1}$

13: **else**

14:     $\alpha_{k+1} = 1, \hat{w}_{k+1} = w_{k-1}, c_k \leftarrow \delta^{-1} c_{k-1}$

15: **end if**

16: **end for**

In the line 10 of the algorithm, we judge whether the residual has reduced by $\delta$ in the most recent step. If so, execution acceleration method. Otherwise, a restart by resetting $\alpha_{k+1}$ to 1 is performed. In order to avoid the acceleration algorithm from being restarted frequently, $\delta$ are suggested to set as 0.99.

## IV. NUMERICAL TESTS

Proposed parallel planning model and solution procedure is tested on planning of backbone grid and different number of sub-areas. System data can be accessed from [40]. All numerical tests were carried out on a laptop with an Intel Core i7-10875H CPU and 16 GB RAM. The MILP model was solved by Gurobi (ver. 9.5.0) [41].

### A. Backbone grid and two sub-areas

The backbone grid consists of eight nodes and each sub-area consists of 20 nodes. Branch and node parameters partly come from [31]. The topology of the test system can be found in the supplementary file [42].

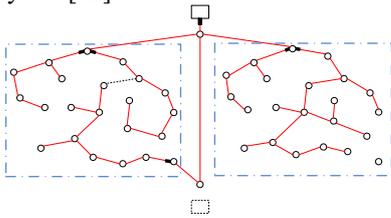

Fig. 3 Extended planning results of the centralized method and proposed parallel computing method

TABLE I
RELIABILITY INDEX AND CONSTRUCTION COST OF CENTRALIZED METHOD AND PARALLEL COMPUTING METHOD

|  | Centralized method | Parallel computing method |
|---|---|---|
| SAIDI Requirement (Area1) |  | 1.5 |
| SAIDI Evaluated (Area1) |  | 1.4997 |
| SAIDI Requirement (Area2) |  | 2 |
| SAIDI Evaluated (Area2) |  | 1.9780 |
| Total cost (10³$) |  | 1007 |
| Solution time (s) | 928.9 | 1335.3 |

In order to verify the performance of the proposed method, the results and solution time of the proposed method was compared with the centralized method [31]. The results of the two planning methods are shown in Fig. 3. Reliability evaluation [5] is used to verify the results to ensure that the results meet the reliability requirements. The red solid line represents the branch which put into operation under normal conditions, and the black dotted line represents the tie line for the purpose of isolation and transfer after a fault. The reliability index and construction cost of the planning results of different methods are listed in the Table I.

It can be seen that the constructed branches and the operation mode planned by the proposed method is the same as the centralized method. The construction cost of the two planning methods is the same. However, since of the parallel solution architecture, the proposed method is faster than the centralized method.

### B. Backbone grid and three sub-areas

Another numerical test is conducted on a larger case with a backbone grid and three sub-areas. The detailed setting of this case can be found in the supplementary file [42].

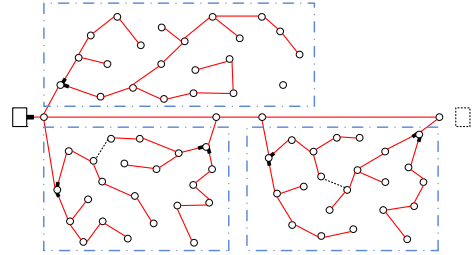

Fig. 4 Extended planning results of the centralized method and proposed parallel computing method

The planning results of the centralized planning method and the parallel computing method are shown in the Fig. 4. Construction cost and solution time of the planning results of the two methods are listed in Table II.

TABLE II
CONSTRUCTION COST AND SOLUTION TIME OF THE PLANNING RESULTS OF THE TWO METHODS

|  | Total cost (10³$) | Solution time (s) |
|---|---|---|
| Centralized method | 1243 | 6980.4 |
| Parallel computing method | 1243 | 1934.6 |

The effect of the acceleration algorithm is shown in Fig. 5. It can be seen that the convergence speed of the method is significantly improved by introducing the acceleration algorithm. The method with acceleration algorithm converges to the optimal value (1243k$) in the 30th iteration, while the method without acceleration algorithm only reach 1255k$ in the same iteration step.

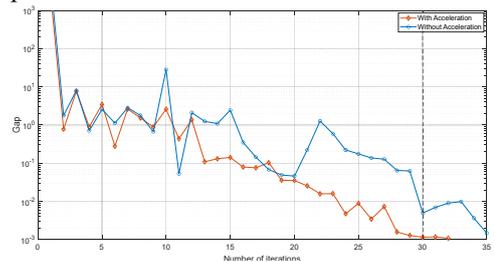

Fig. 5 Comparison of the effects of acceleration method

The effect of the proposed method on multi-stage planning problems and the model considering the uncertainties of load and DG are shown in the supplementary material [42].

## C. Backbone grid and six sub-areas

We further expand the scale of the case to a 139-node DN expansion planning problem containing a backbone grid and six sub-areas. Because the scale of the centralized model is too large, it cannot be solved in an acceptable time. There are only the results of the parallel computing method shown in the Fig. 6. The reliability index and construction cost of the planning results are listed in the Table III. When the centralized method is adopted, the gap of solution is still 22.02% after seven days, and the total cost of the planning scheme is 51333 k$.

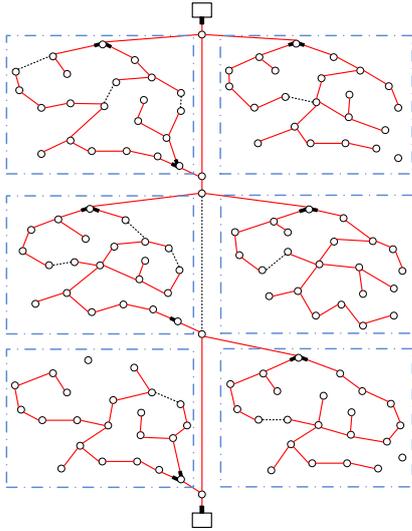

Fig. 6 Extended planning results of the proposed parallel computing method

TABLE III
RESULTS OF PARALLEL COMPUTING METHOD WITH ACCELERATION FOR 139-NODE SYSTEM

|  | Total cost ($10^3$\$) | SAIDI requirements | SAIDI | Solution time (s) |
| --- | --- | --- | --- | --- |
| Parallel computing method | 40429 | 1.9 | 1.8481 | 5358.3 |
|  |  | 2.6 | 2.5951 |  |
|  |  | 2.2 | 1.9973 |  |
|  |  | 2.8 | 2.7961 |  |
|  |  | 2.4 | 2.3895 |  |
|  |  | 2.9 | 2.7823 |  |

## D. Large-scale case

To further verify the scalability of the proposed method, a large-scale network consists of a backbone grid and 10 sub-areas is used to test the proposed method. The case is stitched together from the systems existing in the literature The backbone grid is a modified 85-node distribution network and the sub-area is a modified 141-node distribution network, the information of which can be found in MATPOWER [43]. The results of the parallel planning method with acceleration are shown in Table IV.

TABLE IV
RESULTS OF PARALLEL COMPUTING METHOD WITH ACCELERATION FOR LARGE-SCALE CASE

| Total cost ($10^3$\$) | SAIDI requirements | SAIDI | Solution time (s) |
| --- | --- | --- | --- |
| 71411 | 1.5 | 1.4977 | 98035.59 |
|  | 1.4 | 1.3494 |  |
|  | 1.7 | 1.6998 |  |
|  | 1.6 | 1.5794 |  |
|  | 1.8 | 1.7983 |  |
|  | 1.5 | 1.4850 |  |
|  | 1.4 | 1.3749 |  |
|  | 1.7 | 1.6699 |  |
|  | 2 | 1.9158 |  |
|  | 1.9 | 1.8987 |  |

## E. Calculation efficiency analysis

It can be seen that the number of binary variables of the centralized method is roughly quadratic with the size of the DN. With the expansion of the size, the solution time will increase dramatically. But for the parallel computing method, the number of binary variables increases linearly with the size of the problem, which has obvious advantages in larger-scale systems planning. We select five cases for comparing, the sub-area of which varies from 2 to 6. The result is shown in Fig. 7.

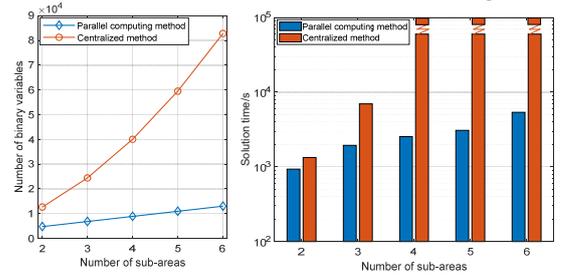

Fig. 7 Comparison of centralized method and parallel computing method in terms of the number of total binary variables and the solution time

It can be seen that as the number of sub-regions increases, the solution time of the centralized method increases dramatically. When the calculation example increases to 4 sub-regions, the centralized method has exceeded 60000 seconds while the proposed method converges within the acceptable time in the all cases.

## V. CONCLUSION

We propose a parallel computing based acceleration method for solving the reliability-constrained DN planning problem. A decomposition planning model containing backbone grid and sub-areas is presented, in which the integer variables increase linearly with the size of the planning networks, while those in the original model increase quadratically. A parallelizable augmented Lagrangian algorithm incorporating Nesterov acceleration algorithm with restart techniques is adopted to solve the decoupling planning model. The numerical tests on different scale cases demonstrate that the proposed method has significant advantages in terms of computational efficiency under the premise of ensuring optimality. The proposed method makes the reliability-constrained DN planning model has the potentiality to be applied in real-world problems.